\pdfoutput=1

\documentclass[11pt]{article}

\usepackage[]{acl}

\usepackage{times}
\usepackage{latexsym}

\usepackage[T1]{fontenc}

\usepackage[utf8]{inputenc}

\usepackage{microtype}

\usepackage{multirow}
\usepackage{graphicx}
\usepackage{subcaption}

\usepackage{xspace}

\newcommand{\sys}{UINav\xspace}
\newcommand{\id}[1]{{\sf\small #1}}

\iftrue
\newcommand{\hide}[1]{}
\else
\newcommand{\hide}[1]{#1}
\fi

\title{UINav: A Practical Approach to Train On-Device Automation Agents}

\author{
 Wei Li\footnotemark[2] \hspace{0.6ex} Fu-Lin Hsu\footnotemark[3]\thanks{Work done as an intern at Google Research.} \hspace{0.6ex} Will Bishop\footnotemark[2]  \hspace{0.6ex}Folawiyo Campbell-Ajala\footnotemark[2] \hspace{0.6ex}
 Max Lin\footnotemark[3]  \hspace{0.6ex} Oriana Riva\footnotemark[2]\vspace{2ex}\\
\footnotemark[2]~\,Google Research\thanks{Now Google DeepMind}\\
\footnotemark[3]~\,Google Cloud
}

\begin{document}
\maketitle

\begin{abstract}
Automation systems that can autonomously drive application user interfaces to complete user tasks are of great benefit, especially when users are situationally or permanently impaired. Prior automation systems do not produce generalizable models while AI-based automation agents work reliably only in simple, hand-crafted applications or incur high computation costs. We propose \emph{\sys}, a demonstration-based approach to train automation agents that fit mobile devices, yet achieving high success rates with modest numbers of demonstrations. To reduce the demonstration overhead, \sys uses a referee model that provides users with immediate feedback on tasks where the agent fails, and automatically augments human demonstrations to increase diversity in training data. Our evaluation shows that with only 10 demonstrations \sys can achieve 70\% accuracy, and that with enough demonstrations it can surpass 90\% accuracy. 
\end{abstract}

\section{Introduction}

The next frontier in artificial intelligence is agents that autonomously operate computers as humans do. Instructed by users in natural language, these agents are especially valuable when their users have visual or motor disabilities or when they are situationally impaired (e.g., driving, cooking). We are particularly interested in agents that can execute human tasks by interacting directly with the user interface (UI) of a running application. These so-called \emph{UI automation agents}~\cite{web-workflows18,Seq2Act:li-acl20,humphreys2022datadriven} can scale well to support a myriad of tasks because they do not depend on third-party APIs. 


Existing approaches to UI automation range from UI scripting to AI-based agents. UI scripts can work reliably, but they involve coding or manual demonstrations~\cite{selenium,ringer16,ETNA,sigilite17} and they cannot tolerate well changes in the UI and workflows, thus leading to high maintenance costs -- this is, however, what enterprises use to automate business workflows~\cite{ui-path}. AI-based approaches can scale better. Using imitation learning and reinforcement learning~\cite{web-workflows18, learning_to_navigate_web}, agents are trained to navigate the web autonomously. However, their synthetic and simplified test environments~\cite{miniwob} and their dependency on large amounts of demonstrations~\cite{humphreys2022datadriven} make them hard to deploy. Recent work leverages Transformers~\cite{Seq2Act:li-acl20,spotlight,venkatesh2022ugif,wang2023conversation} and pre-trained large language models (LLMs)~\cite{yan2023gpt4v,venkatesh2022ugif,zheng2023seeact}. Despite the performance improvement, these solutions come with large resource costs (multiple days of training on hundreds of GPUs/TPUs and high inference costs).

A practical approach to UI automation requires trading between accuracy, generalizability and computational costs. We find a sweet spot between these three properties, and propose \emph{\sys}, a demonstration-based system designed to produce lightweight neural agents that can run on mobile devices while yielding good success rates. 

As in prior work, \sys needs to address the challenge of how to achieve good success rates with fewer demonstrations. We observe that the demonstrations required to achieve good performance differs widely across tasks and environments. If the environment is relatively static even a handful of demonstrations is sufficient; for tasks that must work across many different UIs more demonstrations are needed. When collecting demonstrations, \sys provide users with immediate feedback on which tasks are failing and may benefit from additional demonstrations, and which are satisfactory. It does so through a \emph{referee} model which is trained with the same set of demonstrations used to train the automation agent, but with a different goal: predicting whether a task is successfully completed (rather than predicting which UI action to perform). 

Another challenge \sys addresses is how to increase the robustness of automation agents against system delays and changes in the UI. It does so through three key techniques. First, every UI action is executed as a small program composed of lower-level operations with status checks. These programs, referred to as \emph{macro action}s, abstract the system-specific details thus greatly reducing the agent's state space and therefore the number of required demonstrations. Second, \sys adopts \emph{demonstration augmentation} where human demonstrations are augmented by randomizing non-critical UI elements to increase their diversity. Finally, through \emph{utterance masking} variable sub-strings in utterances are abstracted out. 

We develop \sys using an internal dataset of 40+ tasks and test is on actual Android phones. We also evaluate it on a public dataset, where \sys outperforms various baselines and demonstrates generalizability. Overall, we make the following contributions: (i) a practical system to build UI automation agents that achieve near perfect success rates on previously seen tasks and that can be deployed to mobile devices; (ii) an error-driven process to collect demonstrations paired with augmentation techniques and macro actions; and (iii) a comprehensive evaluation demonstrating \sys's advantages over state-of-the-art systems. 

\section{Related work}

\paragraph{UI automation scripts.}
Record-and-replay tools like Selenium~\cite{selenium} can be used to facilitate the generation of UI automation scripts. These scripts can also be integrated with robotic process automation tools~\cite{ui-path,automation-anywhere,blue-prism}. Programming by demonstration tools~\cite{scrapbook98,Leshed08,vegemite09,Li10,ringer16, sigilite17, rousillon18} are advanced record-and-replay tools that can generate fully functional UI scripts and even action graphs~\cite{ETNA} from recordings of task interactions (demonstrations), which could also be provided in the format of video recordings~\cite{videorpa-icse20,chen2022extracting}. Overall, a major downside of this line of work is that these systems do not produce models that generalize to new task workflows and UIs. 


\paragraph{AI-based automation.}
Transformer-based architectures~\cite{Seq2Act:li-acl20,bai2021uibert,action-bert21,lexi22,spotlight} and reinforcement learning approaches~\cite{web-workflows18,learning_to_navigate_web,glider:sigir21} have been proposed to train agents capable of navigating apps and websites when provided with natural language instructions. Yet, it is unclear how well these systems perform in a variety of real-world environments and scale across task categories because either they have been tested in synthetic webpages of 10--50 UI elements~\cite{miniwob} or on limited datasets~\cite{Seq2Act:li-acl20,motif}. 
Recent work leverages LLMs to ground natural language instructions in UIs~\cite{venkatesh2022ugif,wang2023conversation,yan2023gpt4v,zheng2023seeact,aitw2023}. These approaches come with a large training overhead (e.g., multiple days of training on hundreds of GPUs/TPUs) and a high inference cost which prevents them from running on mobile devices.

In this paper, we extend our previous work~\cite{macro_action} where macro actions were introduced but was limited to work with OCR and icon recognition, into a full system, that bridges the gap between programming by demonstrations and AI-based systems by providing an easy-to-learn system to train robust, multi-task agents for UI navigation in real-world applications. While the system requires manual demonstrations for training, it provides an error-driven collection of demonstrations where testing scenarios are automatically generated and evaluated by the system, thus reducing the number of redundant demonstrations. The error driven demo collection of \sys is inspired by the DAGGER~\cite{pmlr-v15-ross11a} algorithm and we show that it is effective in reducing the number of demonstrations for both sequential (referee) and non-sequential (agent) models.

\section{Why is UI automation hard?}
We study the problem of how a UI automation system can generalize to new execution environments, including different apps and different tasks, \emph{without} requiring an excessive number of demonstrations. To illustrate the challenges we use an apparently simple task, \textit{search}, i.e., operating the search bar of an app. 
Two aspects make this task challenging. 

Search is a universal task that must work across a myriad of apps where search bars can take many different formats. Some search bars require the user to type some keywords and then click an icon (typically on the right hand-side); others, as the user types, automatically display search results which can be directly opened; some others have an additional field (e.g., a category) which must be set beforehand; there are also search bars that are hidden and reveal only upon clicking on an icon; etc. 

The second axis of complexity regards the agent's start state. When an agent is requested to search in a specific app, the user's device screen may not display the target app or may display it in a page (state) without any search functionality. The agent must first understand how to navigate to the state offering the search function, which may involve navigating back, launching a different app, or dismissing welcome screens and ads. Even when the environment already shows the desired search widget, its state may need to be reset, e.g., by deleting search terms previously entered (see the YouTube example in Fig.~\ref{fig:search_youtube} in the Appendix). 

In general, in a real environment, an agent is exposed to many different screen conditions caused by a combination of factors: different apps, dynamic app content, previous interactions, layout variance across devices, UI changes across app/OS versions, ads and notifications, etc. An agent needs to ignore irrelevant UI elements and navigate to relevant states. One way to tackle this variability is through more demonstrations, but with obvious overheads. \sys's first contribution is to adopt an error-driven process to collect demonstrations (\S\ref{sec:referee_model}). Its second contribution is to amplify the learning brought by each demonstration by automated augmentation (\S\ref{sec:demo_augmentation}). Finally, to address variability issues due to system delays, rather than relying on demonstrations \sys takes a programmatic approach by introducing macro actions (\S\ref{sec:macro-actions}).

\section{System design}
\label{sec:system_design}

\begin{figure}[t]
\centering
\includegraphics[width=\columnwidth]{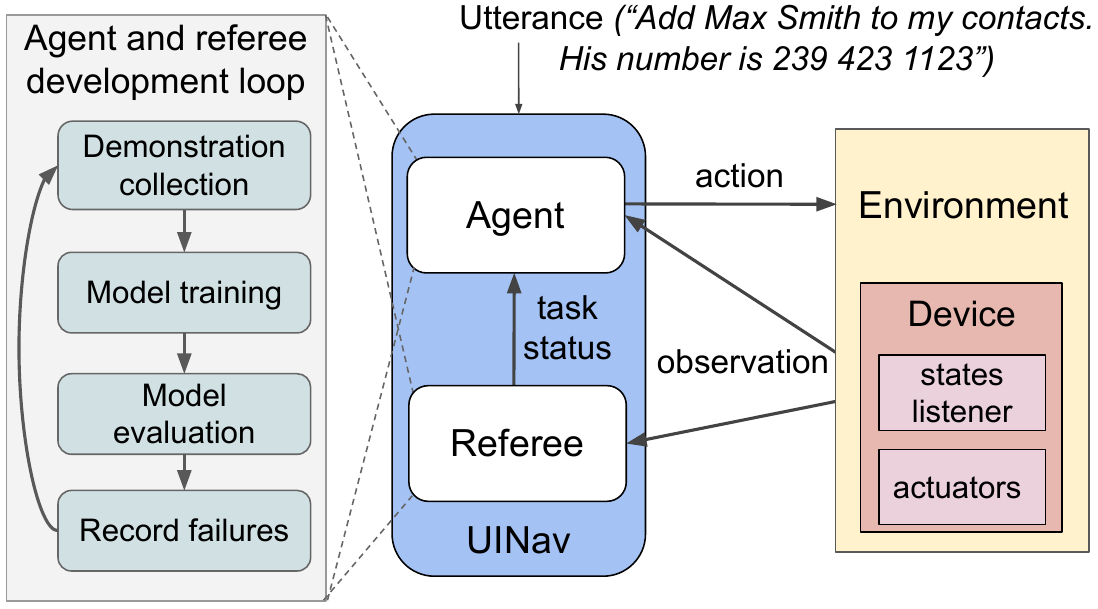} 
\caption{High-level architecture of \sys.}
\label{fig:uinav_development}
\end{figure}

Fig.~\ref{fig:uinav_development} shows the high-level architecture of \sys. Given a task represented by a natural language utterance and an observation of the device state (i.e., a representation of what is currently displayed on the screen), a neural network-backed agent responds with its choice of action to complete the task. The predicted action is executed by the environment by interacting with a device's system (an emulator or a real phone). Then, the agent is provided with a new observation describing the new state and a new action is predicted. This setup is similar to that of a reinforcement learning agent, but \sys also includes a second agent called \emph{referee}, which is responsible for judging the completion status of a task (episode) at each time step. 

The development of \sys agents (left of Fig.~\ref{fig:uinav_development}) involves first collecting human demonstrations for some target tasks, then training the neural networks of the agent (\S\ref{sec:agent_model}) and referee (\S\ref{sec:referee_model}), 
and finally evaluating them on the device. Failures of either the agent or the referee are recorded and used to guide the collection of new demonstrations to be used in the next round of training. The development loops over these steps until no more errors of either the agent or the referee are found.

\subsection{Agent's neural network architecture}
\label{sec:screen_representation}
\label{sec:agent_model}
\begin{figure}[t]
\centering
\includegraphics[width=\columnwidth]{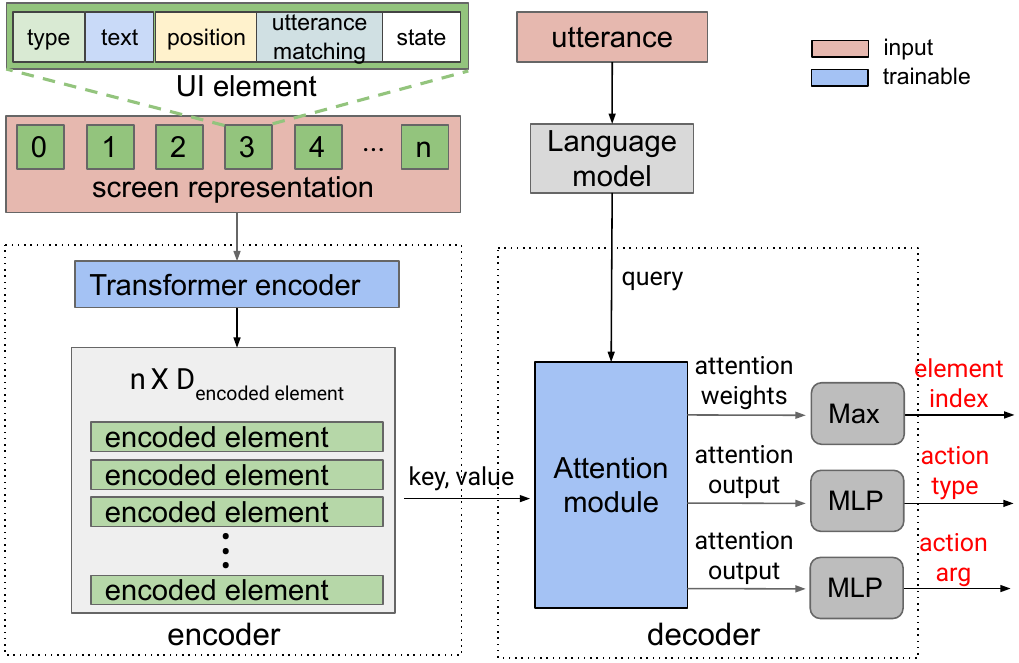} 
\caption{The neural network of the agent model.}
\label{fig:agent_architecture}
\end{figure}

\hide{We now describe the architecture of the agent model (Fig.~\ref{fig:agent_architecture}), and focus particularly on the techniques that enable this architecture to reduce the number of required human demonstrations.}

The \sys agent consists of an encoder-decoder architecture (Fig.~\ref{fig:agent_architecture}). It perceives the state of the device through observations of what is currently displayed on the screen, represented by the set of UI elements composing it. Each UI element is described by a set of attributes: type (button, icon, etc.), text (visible text, content description, resource identifier, etc.), on-screen position, utterance matching (whether on-screen text matches the utterance\footnote{Similarly to previous work~\cite{web-workflows18}, we compute utterance matching as the average of the similarity scores of all words in the UI element's text with the utterance.}), and state (e.g., whether a checkbox is selected). The screen representation can be generated from raw pixels processed by screen understanding techniques~\cite{uied-icse20,screen-parsing21, screen-understanding-apple-chi21}, which also include icon detection and text recognition, or from a tree-structured representation of the UI, such as the Android accessibility tree. Our implementation dynamically switches between the two sources of screen representation based on simple heuristics, such as whether the target app is known to provide poor accessibility support or whether the number of accessibility nodes is extremely small.

Then, the input to the neural network of the agent is a set of UI elements and an utterance. Each UI element is represented by a vector concatenated from the feature vectors of its attributes. Text labels of UI elements are encoded by a language model~\cite{devlin-etal-2019-bert}. The feature vectors of the UI elements are fed into a Transformer encoder. The output of the encoder is a function of the encoding of each UI element plus its attention over all other UI elements on the screen, including itself. 

The decoder predicts which action to perform. This involves predicting \emph{(i)} the UI element on which to perform the action, \emph{(ii)} the type of action (click, type, etc.), and \emph{(iii)} any argument for the action. Actions (summarized in Table~\ref{tbl:action_space}, \S\ref{sec:action_space}) can be of two types. Element actions (\id{click}, \id{focus\_and\_type}, \id{dismiss}) manipulate a specific element, while global actions (\id{wait}, \id{back}, \id{scroll}, \id{open\_app}) are general operations or platform-specific functions. 

The decoder uses a single cross-attention module, with the utterance embedding serving as the query vector and element encodings serving as keys and values. The largest attention weight is used to select the element to act upon, while the vector output of the cross-attention module is passed through two independent multi-layer perceptrons (MLP) to predict action type and argument. 

In its essence, the agent's neural network implements a scoring system. For any given screen, all its elements are scored, and the highest-scored one is selected. Due to the attention in the encoder, for any UI element, its relationship with all the other elements can be encoded. The Transformer model learns how different combinations of UI elements and utterances map to actions, and uses this knowledge to rank elements to act on. It is essential that the model learns to evaluate single UI elements in the \emph{context} of others because the meaning of UI elements is often context sensitive~\cite{lexi22} -- elements of similar appearance (color, size and shape) can have different functions but neighboring elements like text labels can help resolve the ambiguity. For specific examples on how \sys contextually evaluates UI elements see \S\ref{sec:case-studies}.

\subsection{Referee model}
\label{sec:referee_model}

In the agent's action space there is no ``done'' action. This means that the agent does not stop on its own but instead relies on the environment to terminate a task. This is common practice in reinforcement learning. Instead of building task-specific termination logic, we train a \emph{referee} model to predict whether a task is completed at each step and what its outcome is. The referee is trained using the exactly same set of demonstrations as the agent, hence it does not incur extra effort in data collection. However, it also serves a second purpose. 

A well-known challenge in demonstration-based systems is that they can require excessive developer time to collect a sufficient number of demonstrations~\cite{Lau09-PBD-fail} and that it may be difficult to provide samples that are sufficiently different from each other~\cite{MYERS200145,human-mistakes17}. By automatically evaluating the execution of a currently-trained agent and identifying failing tasks, the referee guides users towards collecting new demonstrations \emph{only} for critical scenarios. Failed executions are saved along with all their parameters and passed to the demonstrator.

The neural architecture of the referee model is similar to that of the agent except that it is wrapped in a recurrent neural network to consider the history of actions (see \S\ref{sec:referee_model_appendix} for more details). The referee predicts one out of 4 labels: (1) \id{SUCCESSFUL}: the task is completed successfully; (2) \id{FAILED}: the task has failed or has reached the maximum number of allowed steps; (3) \id{PENDING}: the task is ongoing; or (4) \id{INFEASIBLE}: the task cannot be executed.

\subsection{Utterance masking}
\label{sec:utterance_masking}



\sys's focus is on generalizing to different execution environments without requiring an excessive number of demonstrations. However, another large source of variability is the input instruction provided in natural language. To address this problem, we design \sys agents to learn general task workflows rather than specific task instances. We do so by pre-processing utterances to identify sub-string that represent the variables of a task. For example, in \textit{Search for {\color{red} tiktok} in Google}, \textit{\color{red} tiktok} is the phrase to search for and can be replaced by other keywords. The variable sub-strings are masked and replaced by \emph{placeholders} before being encoded, so that the utterance embedding is independent on the specific instances. As a result, there is no need to train with different utterances covering the distribution of variables.

For any utterance, all the replaced sub-strings are included in the list of entities associated with the task. A matching vector is computed for each UI element on the screen and is included in the element attributes passed as input to the agent. In the matching vector, each scalar is in the range of $[0, 1]$ and computed as the cosine similarity between the text label of the UI element and the corresponding entity string. 

Variable sub-strings can be identified by either following pre-defined patterns, through the use of explicit delimiters, or semantic parsers~\cite{kamath2019survey}. LLMs can also be employed~\cite{shin-van-durme-2022-shot, drozdov2022compositional, mekala2022zerotop}. \sys still works without utterance masking but may require more demonstrations to reach similar accuracy (see ablation analysis in Table~\ref{tab:agent_on_motif}).

\section{Increasing robustness and efficiency}
\label{sec:robustness}
\label{sec:macro-actions}
\label{sec:demo_augmentation}

We have described how \sys helps developers balance accuracy and number of demonstrations. Next, we describe the techniques that increase the agent robustness in the face of system delays, UI changes, and variations in task descriptions.

\paragraph{Action validation and macro actions.} Controlling UIs of an actual device involves dealing with various system issues. There are unavoidable delays between the time a state is collected from a device and when a predicted action is ready to be performed. Screens can also be slow at loading or updating, hence an agent needs to wait for them to stabilize. These delays are particularly noticeable on a mobile device. To deal with these issues, rather than modeling this variability through more demonstrations, we take various programmatic measures. 

First, before executing an action, \sys validates it by checking whether a referenced UI element is still on the current screen and if so, whether it has changed. If the action is not applicable anymore, it requests a new prediction. 

Second, every action is executed as a small program that is composed of lower level operations with status checks. Such a program is referred to as \emph{macro}. Each macro is implemented following a state transition graph and it is atomic so that while a macro is running the agent stays idle and changes to the screen are not visible to it. An example of macro action is \id{focus\_and\_type} which comprises 4 low-level actions: clicking the input field to obtain focus, waiting for the blinking cursor to appear, typing the text in the field, and (optionally) pressing Enter. See \S\ref{sec:macro_actions_details} for more details.

\paragraph{Demonstration augmentation.}
To further limit the number of required demonstrations and amplify the learning brought by each one, \sys also augments the collected demonstrations by randomizing the attributes of randomly-selected, non-critical UI elements. This teaches the agent which elements may be safely ignored, and ultimately makes it more tolerant to UI changes. Non-critical UI elements have their attributes modified with a predefined probability by either (i) replacing the embedding of their text labels with random vectors, or (ii) by adding random offsets to the four scalars of their bounding boxes, which is equivalent to randomizing both the element's position and size. Despite its simplicity, demo augmentation is highly effective at improving \sys's performance (see Table~\ref{tab:agent_on_motif}).

\begin{table}[t]
\centering
 \caption{Inference time (msec) on high/low-end phones. None of the models utilize any accelerators.}
 \scalebox{0.9}{
        \begin{tabular} {llll|l}
            Device & Agent & Referee & SmallBERT & Total\\
            \hline
            High-end &1.98 & 2.21 & 262.79 & 267.00\\
            Low-end & 4.40 &5.24 & 427.63 & 437.27\\
        \end{tabular}
        }
    \label{tab:on_device_execution_time}
\end{table}

\section{System evaluation}
\label{sec:evaluation}

We built \sys for Android. Both the agent and referee are implemented in TensorFlow. The agent model has 320k parameters and its tflite version occupies 1.3MB, while the referee has 430k parameters and it is 1.8MB large. For text encoding we use SmallBERT~\cite{smallbert} and convert it to a 17.6MB tflite model. No quantization is applied during the conversion (More implementation details in \S\ref{sec:impl_details}). As shown in Table~\ref{tab:on_device_execution_time}), both the agent and referee take only a couple of milliseconds to execute on a high-end phone (e.g., Pixel6pro) and around 5 milliseconds on a low-end phone (Pixel 3a). BERT dominates the total time.

\begin{table}
\centering
\caption{Task and step accuracy on MoTIF.} 
\scalebox{0.75}{
    \begin{tabular} {lcccc}
       Model & \multicolumn{2}{c}{App seen task unseen} & \multicolumn{2}{c}{App unseen task seen} \\
        & task acc & step acc & task acc & step acc \\
            \hline
            Seq2Seq & 22.5\% & 40.4\% & 18.0\% & 31.3\%\\
            
             MOCA & 21.3\% & 40.0\% & 17.0\% & 32.7\%  \\   
            
             Seq2Act & 32.4\% & 66.4\% & 28.3\% & 67.7\%\\
            
            \sys & 37.9\% & 73.7\% & 36.8\% & 66.8\%\\
           
            UINav+aug & 39.4\% & 74.9\% & 39.7\% & 68.4\%\\   
           
            UINav+aug+utt  & \textbf{68.3\%} &\textbf{89.7\%} & \textbf{59.6\%}& \textbf{81.9\%}\\
            \hline
        \end{tabular}
        }
    \label{tab:agent_on_motif}
\end{table}


\subsection{Agent and referee accuracy}
We evaluate \sys on the MoTIF dataset~\cite{motif}. MoTIF includes two splits: (i) \emph{app seen task unseen} which tests whether a model can generalize to new tasks, and (ii) \emph{app unseen task seen} which tests whether a model can generalize to new apps. As in the evaluation of the MoTIF system, we train \sys using low-level instructions, and compare against three baselines: Seq2Seq \cite{Seq2Seq}, MOCA~\cite{MOCA}, and Seq2Act~\cite{Seq2Act:li-acl20}. More training details in \S\ref{sec:training_details}.
We measure \emph{(i) step accuracy}, the percentage of task steps where the model and the dataset have matching outputs, and \emph{(ii) task accuracy}, the percentage of tasks with all steps matching.


Table~\ref{tab:agent_on_motif} reports the results. For ablation purposes, we consider three variants of \sys, depending on whether demonstration augmentation (+aug) and utterance masking (+utt) are enabled. UINav+aug surpasses all baselines by 7 and 11 percentage points in task accuracy and 8.5 and 0.7 points in step accuracy. Without demo augmentation \sys outperforms all three baselines, in all except one case (step accuracy in app unseen and task seen). This demonstrates the effectiveness of the \sys design and how demo augmentation effectively exposes the model to a larger variety of training conditions thus improving generalizability. In this dataset, generalizing to new apps appears to be harder than generalizing to new tasks. With the addition of utterance matching, on unseen apps, \sys still achieves 59.6\% in task accuracy and 81.9\% in step accuracy, well above all baselines.

\begin{figure}[t!]
\centering
\includegraphics[width=0.88\columnwidth]{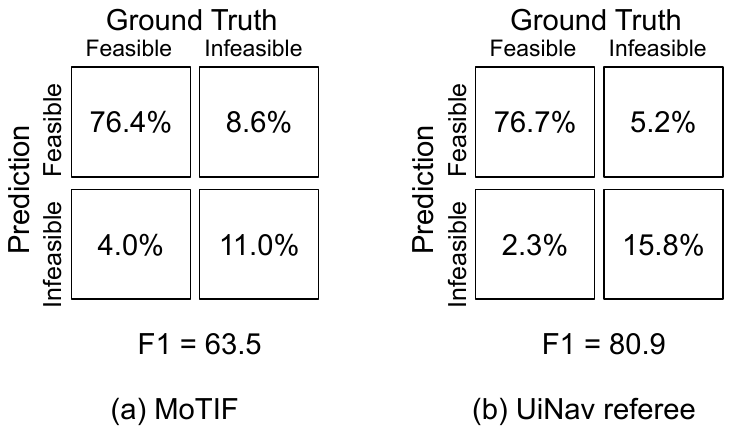}
\caption{Referee model compared to the MoTIF system~\cite{motif} using the MoTIF dataset.} 
\label{fig:referee_vs_motif}
\end{figure}

To evaluate the referee model we use again the MoTIF dataset as its traces are labeled as ``feasible'' or ``infeasible'', depending on whether the task was successfully completed. We compare against the MoTIF system, specifically designed to predict task feasibility/infeasibility. As the \sys referee predicts 4 states, we map \texttt{SUCCESSFUL}/ \texttt{PENDING} to ``feasible'' and \texttt{FAILED}/\texttt{INFEASIBLE} to ``infeasible''. As Fig.~\ref{fig:referee_vs_motif} shows, our referee model produces a significantly better F1 score, 80.9\% vs. 63.5\%, and it is especially better in identifying infeasible tasks.

\subsection{Demonstration effort}

\begin{figure*}[t]
\centering
\includegraphics[width=\textwidth]{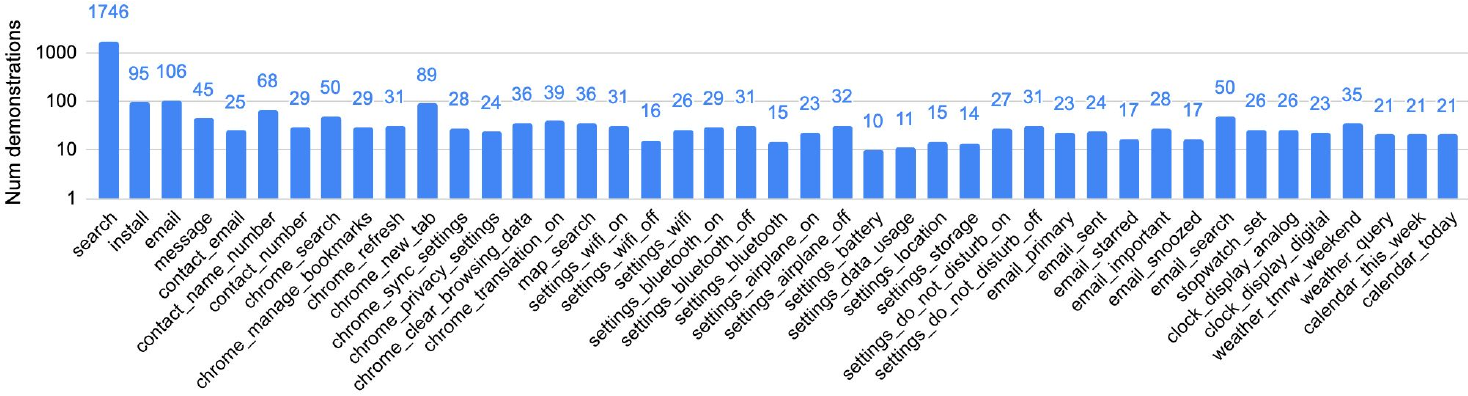} 
\caption{Number of demonstrations in the training set collected for 43 tasks across 128 apps/websites.}
\label{fig:num_training_episodes}
\end{figure*}

To evaluate the effectiveness of the error-driven demo collection approach of \sys we cannot use static datasets. Hence, we quantify the demonstration effort of \sys by using it to train high-accuracy agents for 43 different tasks across 128 Android apps and websites, selected based on popularity (e.g., Gmail, Contacts, Amazon, Airbnb, linkedin.com, target.com, etc.). Please see \S\ref{sec:apps_websites} for the full list. For demo collection we build a dedicated GUI which can be connected to Android phones or emulators (see \S\ref{sec_appendix:ui_nav_console}). The GUI supports macro actions and error-driven data collection. During data collection and testing, the environment automatically performs a few random operations at the beginning of each task, including randomly changing pixel densities, font scales, device orientation, and issuing a sequence of random number of clicks on randomly selected UI elements. The purpose is to start a task from a random state and to diversify data coverage. 


We collect demonstrations with the goal to achieve near perfect success rates. With the exception of the \id{search} task we collect from 10 to 106 demonstrations (on average 32.7) per task, 3661 in total (Fig.~\ref{fig:num_training_episodes}). Collecting 10 demonstrations takes less than 10 minutes. The \id{search} task must work across 100+ apps hence requiring 1700+ samples. To verify this data is sufficient to train accurate agents, in a second phase we collect additional 596 test samples. Because of the random initialization of the environment, and the dynamic characteristics of a live system, it is unlikely that the models see a training sample that is identical to a test one. The \sys agent achieves 90.6\% task accuracy and 95.8\% step accuracy; the referee is 99.5\% accurate. Please note that the numbers of demonstrations in Fig.~\ref{fig:num_training_episodes} are most likely more than the minimum required to reach the same accuracy, as we prioritize improving accuracy over reducing the number of training samples. It is less effort to add new demonstrations as a batch than finding out whether a specific demonstration improves model accuracy.

In an informal user study, a few software engineers with no prior experience using \sys utilized it to build agents for a few tasks. They started from scratch, without using any existing demonstrations. The time spent on collecting data for each task was between 10 to 20 minutes while all participants claimed their resulting agents performed perfectly.

\subsection{Multi-task vs. single-task agents}

\begin{figure}[t]
\centering
\includegraphics[width=\columnwidth]{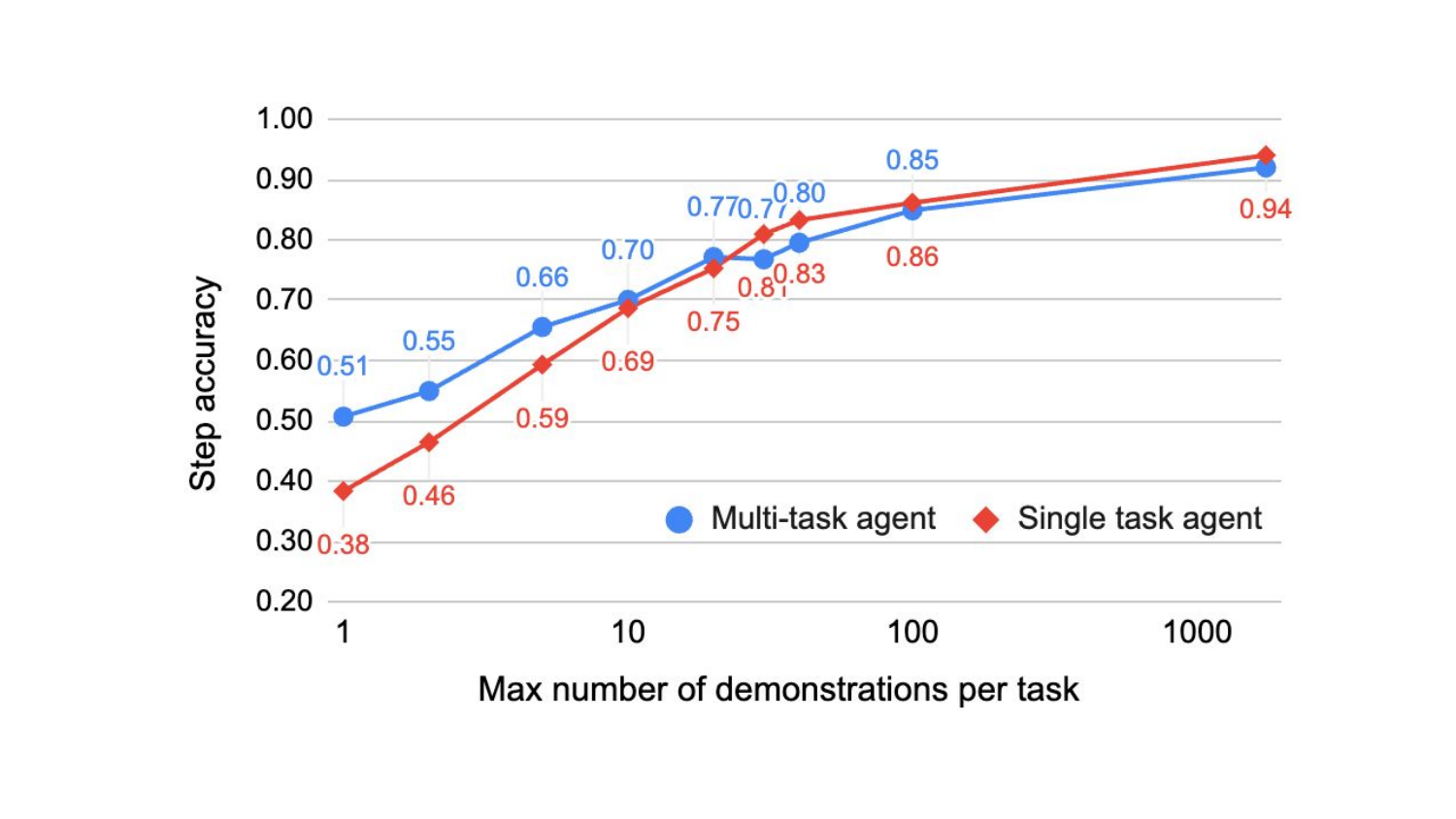} 
\caption{Comparison between multi- and single-task agents with an increasing number of demonstrations.}
\label{fig:episodes_per_task}
\end{figure}

To reduce the resource overhead on mobile devices, we train a single multi-task agent. We show this choice is preferable also for small numbers of demonstrations. From our in-house dataset, we select the 10 tasks with the largest number of demonstrations. We then train one multi-task \sys agent using demonstrations across all 10 tasks and 10 single-task \sys agents using demonstrations from individual tasks. We repeat the training for an increasing number of demonstrations. As Fig.~\ref{fig:episodes_per_task} shows, the multi-task agent reaches 51\% accuracy even with just one demonstration, demonstrating transfer learning across tasks is happening. The average accuracy for both multi-task and single-task agents surpasses 80\% with 40 demonstrations. 

\section{Limitations}

To limit the number of required demonstrations, the \sys agent makes decisions based only on the contents of the current screen and does not utilize information from previous screens. However, if a task truly requires an agent to remember previous states or actions, then the current architecture of the agent model will fail. Our assumption is that a well-designed UI often presents all the information that is needed for successful human interaction on the current screen. The accuracy of our memory-less agents proves that this is the case for the tasks tested so far. For tasks or UIs that require memory, the \sys agent model can be enhanced with memory through either a recurrent neural network or by padding previous states in its input.

Our approach depends on UI elements for both representing features of screens as well as defining actions. It will not work if a screen representation fails to capture critical UI elements. This can happen also when accessibility trees miss critical nodes because content embedded in WebViews and Canvas is generally not captured.  

\section{Conclusions}
We presented a demonstration-based system for building small and fast UI automation agents that are suitable for mobile devices. Our approach requires small human effort and no coding skills. With modest numbers of demonstrations \sys agents achieve near perfect success rate on previously-seen tasks and with more effort they can generalize well to new tasks and applications.

\bibliography{references,misc_references}

\newpage

\appendix
\section{Appendix}

\section*{Ethical considerations}

A use case that motivates \sys agents include screen readers for visually-impaired users. As accessibility labels are often missing or incomplete in mobile apps, \sys can provide them with access to a much wider range of applications and functionality. Another potential use case of \sys is task automation, which has societal, security and privacy implications. An agent may leak private information or carry out a task in an unacceptable way or produce unwanted side effects. Malicious actors could also use \sys agents for undesired purposes such as overriding anti-fraud mechanisms or manipulating applications to achieve undesirable goals.

To develop \sys we collected a dataset internally. The demonstrators were asked to avoid entering any private information and received fair compensation. 

\begin{figure*}[t]
    \centering
    \subfloat[]{\includegraphics[height=0.4\linewidth]{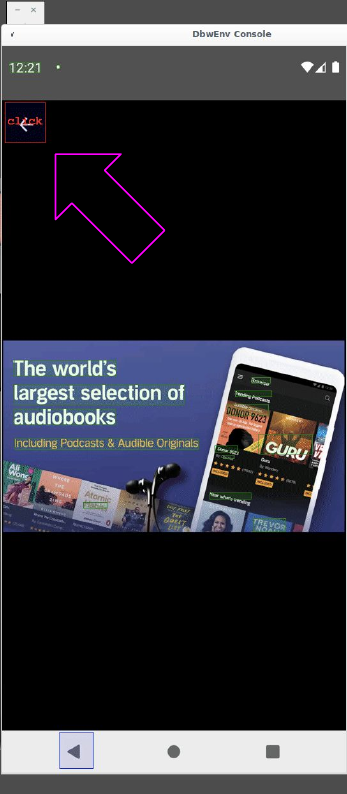}}
    \subfloat[]{\includegraphics[height=0.4\linewidth]{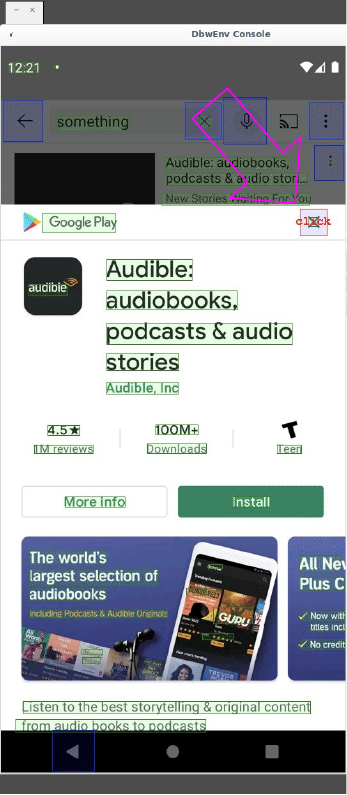}}
    \subfloat[]{\includegraphics[height=0.4\linewidth]{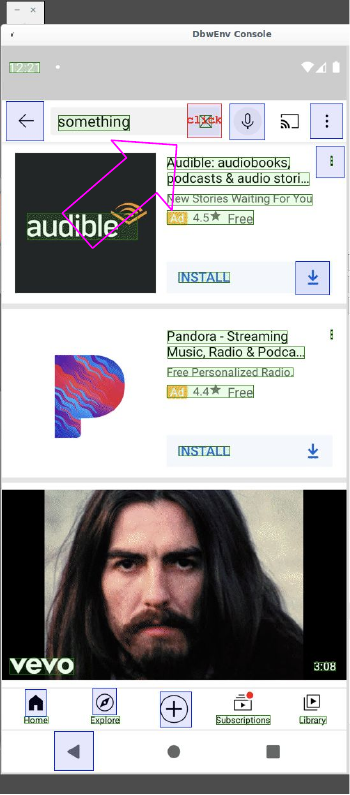}}
    \subfloat[]{\includegraphics[height=0.4\linewidth]{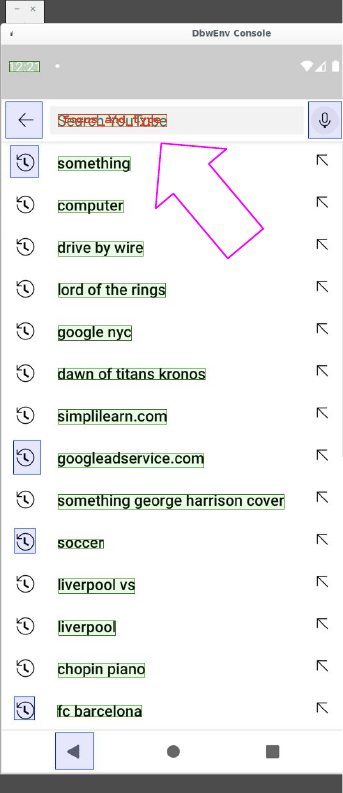}}
    \caption{\sys agent searches in YouTube. The pink arrows highlight the agent's actions that are also annotated by red boxes and texts. To start using the search bar the agent must first dismiss popups (twice) and clear the search bar. (a) Clicks the back button to dismiss a popup ads; (b) Clicks "X" to dismiss the install page of Audible; (c) Clicks "X" to erase the previously entered search phrase ``something''; (d) Focuses on the search bar to enter a new search term.}
    \label{fig:search_youtube}
\end{figure*}

\subsection{An example task: search in YouTube}
\label{sec:num_training_episodes}

Fig.~\ref{fig:search_youtube} shows the \sys agent searching in YouTube. The agent dismisses popups twice (a) and (b) to reveal the search bar. It then clicks the "X" button to erase the previous search phrase ``something'' (c). The system does not reach the desired start state for a search until the screen shown in (d), where the agent sets the focus on the search bar to then enter the search term.

Fig.~\ref{fig:num_training_episodes} shows the \texttt{SEARCH} task requires over 1700 task demonstrations because it must work for 100 or more different apps and websites. All other tasks are specific to a single app and thus require fewer samples, 33 on average.

\subsection{Action space}
\label{sec:action_space}

The types of action the agent can predict define its action space, summarized in Table~\ref{tbl:action_space}. 
Actions can be of two categories. Element actions manipulate a specific element. Global actions are general operations or wrappers for platform-specific functions (e.g., for launching an app). All the tasks that we have tested so far are solvable by these two categories of actions. In the future, we expect to expand the action space to incorporate additional functionality including deep-links and APIs. 

\begin{table*}[t]
\centering
\caption{\sys action space.}
\scalebox{0.9}{
\begin{tabular}{llp{7cm}}
\hline
\multirow{3}{*}{Element} & \id{click <elem>}  & Clicks the center of the specified element.\\
 \multirow{3}{*}{actions} & \id{focus\_and\_type <elem,text>} & Sets focus on the specified element, types the specified text, and optionally presses Enter. \\
  & \id{dismiss <elem>} & Clicks outside of the specified element.\\
\hline
\multirow{4}{*}{Global} & \id{wait} & Waits until the next observation is received.\\
\multirow{4}{*}{actions}  & \id{back} & Goes back to the previous app screen. \\
                         & \id{scroll <left|right|up|down>} & Scrolls in the specified direction.\\
                        & \id{open\_app <app\_name>} & Launches the specified application.\\
\hline
\end{tabular}
}
\label{tbl:action_space}
\end{table*}

\subsection{Referee model}
\label{sec:referee_model_appendix}



\begin{figure}[ht]
\centering
\includegraphics[width=\columnwidth]{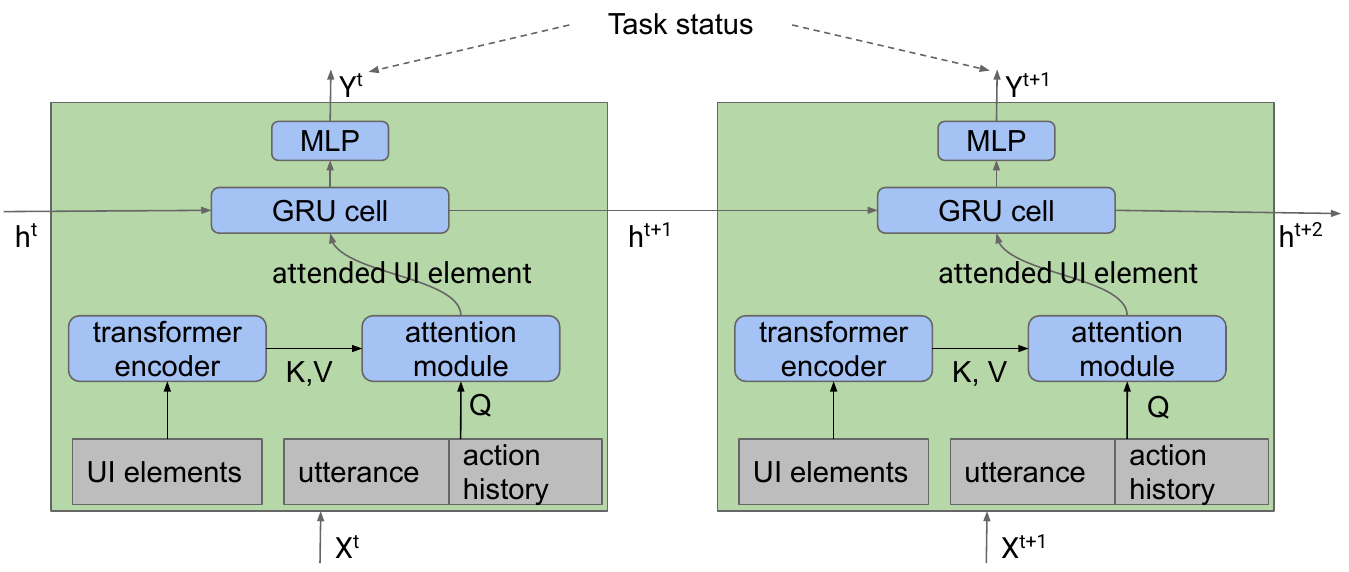} 
\caption{The architecture of the \sys referee model.}
\label{fig:referee_architecture}
\end{figure}

The referee is a recurrent neural network (RNN)-based model (Fig.~\ref{fig:referee_architecture}). The attention over Transformer-encoded UI elements is similar to that of the agent model, except that the query is the input utterance concatenated with the action history (the action performed in the previous step and its outcome). Although action history could be derived from previous screen representations, feeding it as input directly makes it less challenging as the referee does not have to learn it. The output of the attention module is then fed into a gated recurrent unit (GRU)~\cite{GRU}. The GRU takes this along with the previous internal hidden state as inputs to predict the current status of the step: (1) \texttt{SUCCESSFUL}: the task is completed and it is successful; (2) \texttt{FAILED}: the task has failed or has reached the maximum number of allowed steps; (3) \texttt{PENDING}: the task is ongoing; or (4) \texttt{INFEASIBLE}: the task cannot be executed (e.g., the task may not be well defined). Failed executions are saved along with all their parameters and passed to the demonstrator. 

\subsection{Macro actions} \label{sec:macro_actions_details}
\begin{figure}
\centering
\includegraphics[width=\columnwidth]{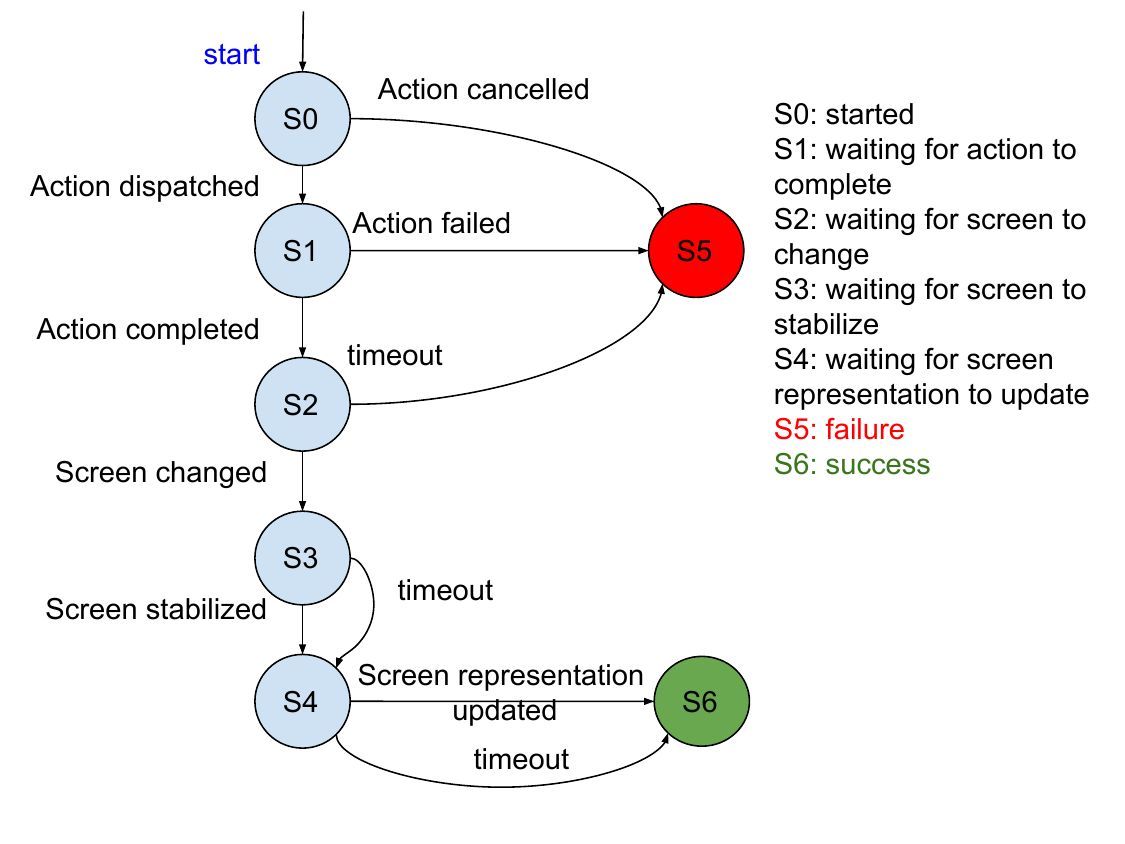} 
\caption{The state transition graph for macro actions resulting in screen changes.}
\label{fig:macro_state_graph}
\end{figure}

In \sys, every action is executed as a small program that is composed of lower level operations with status checks. Such a program is referred to as \emph{macro}. Macro actions abstract the system-specific details, thus making it possible to build cross-platform agents and simplifying the agent's logic. Each macro action is implemented following a state transition graph. Fig.~\ref{fig:macro_state_graph} shows the state transition graph for most macro actions that result in screen changes, such as \texttt{click} and \texttt{back}. It starts at S0, and transitions among the other states according to incoming events, such as \textit{Action dispatched} and \textit{Screen changed}, and exits either successfully (S6) or with a failure (S5). The graphs of other macro actions are similar.

Each macro is atomic so that the agent stays idle while a macro is running. During the execution of a macro action, changes to the screen are not visible to the agent, and do not contribute to the state space. In particular, each macro action is designed to encapsulate transitional screens, and finishes when the screen becomes stable or a timeout is reached (required for dynamic screens such as playing a video).

\begin{figure*}
    \centering
    \subfloat[]{\includegraphics[width=0.2\linewidth]{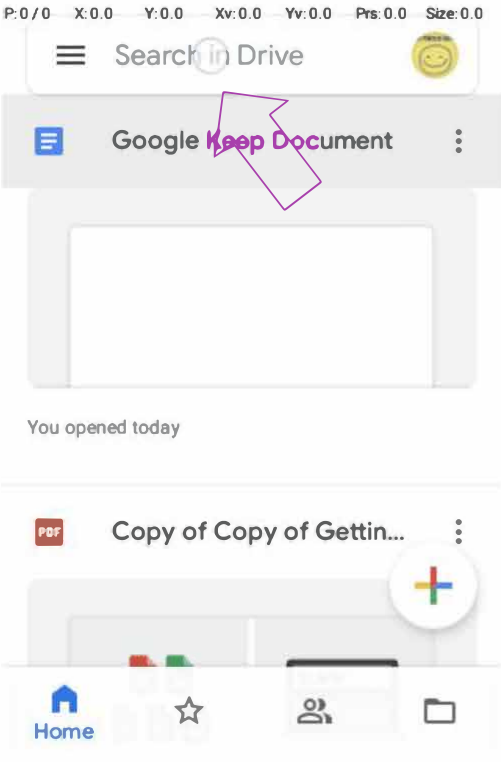}}
    \subfloat[]{\includegraphics[width=0.2\linewidth]{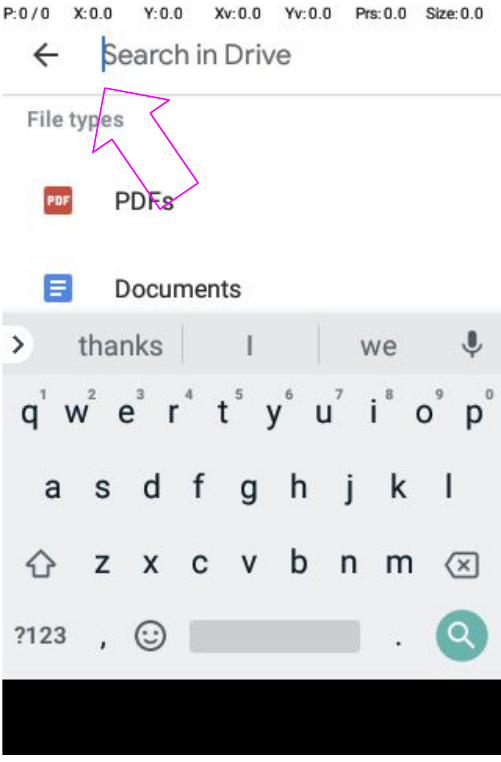}}
    \subfloat[]{\includegraphics[width=0.2\linewidth]{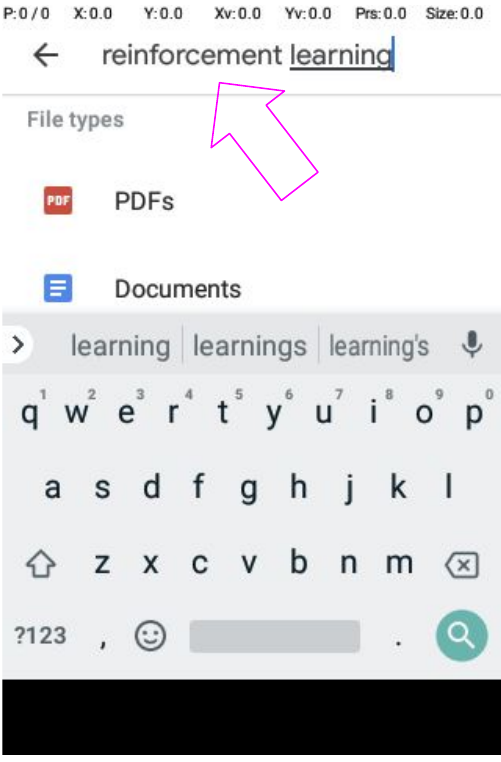}}
    \subfloat[]{\includegraphics[width=0.2\linewidth]{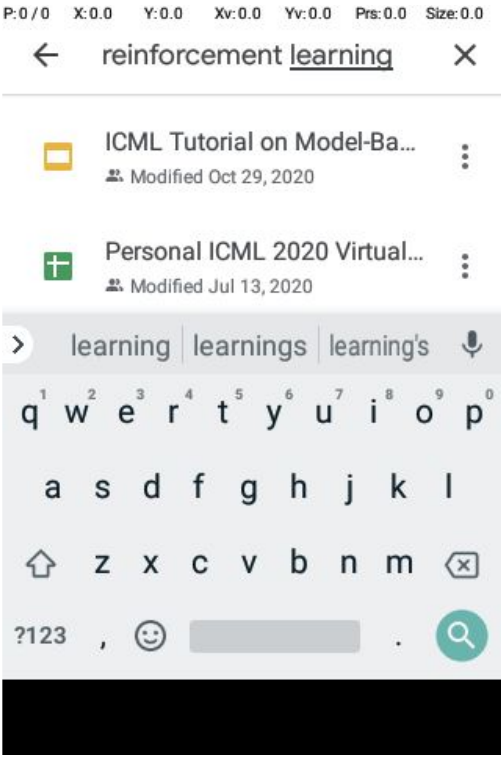}}
    \caption{The \texttt{focus\_and\_type} macro action consists of four steps: (a) clicking the input field (“Search in Drive”) to obtain focus; (b) waiting for the blinking cursor to appear; (c) typing the specified text (``reinforcement learning''); and (d) pressing Enter and wait for the screen to update.}
    \label{fig:focus_and_type}
\end{figure*}

Another advantage of using macro actions is that they package highly dependent, low-level actions. Fig.~\ref{fig:focus_and_type} shows an example. The \id{focus\_and\_type} action (inspired from MiniWoB~\cite{miniwob}) consists of 4 low-level actions: clicking the input field to obtain focus, waiting for the blinking cursor to appear, typing the text in the field, and (optionally) pressing Enter. (Note that large arrows in purple are drawn to highlight interesting areas.) 

As a result, we are able to utilize a memory-less neural network architecture for the agent. In other words, our agent picks an action based only on the information of the current screen. This makes the neural network easier to train. Additionally, a memory-less neural network can be trained using sets of single screenshots, rather than long sequences of screens which can be hard to collect.  

\subsection{Implementation}
\label{sec:impl_details}

We built \sys for the Android platform. However, our design is applicable to other platforms and some of our techniques (e.g., macro-actions and screen representation) are specifically designed to be platform agnostic. Both the agent and the referee models are implemented in TensorFlow. We employ two inference modes, off-device and on-device. During development we use the Python API of TensorFlow to test the models off-device. Once stable, the models are converted to TensorFlow Lite (tflite) for on-device inference. Both agent and referee models utilize the same pre-trained language model to encode utterances and texts appearing on screens. We choose the smallest model, L-2\_H-128\_A-2, of SmallBERT~\cite{smallbert}, and convert it to a 17.6MB tflite model. Note that no quantization is applied during the tflite conversion of any of the above models. For efficiency, the sentence encoding computation of the agent and referee models are shared. 

The selection of SmallBERT over a larger language model is mainly for on-device inference. We restrict the input utterances to predefined patterns so that arguments can be parsed through regular expressions. With the help of utterance masking, our models deal with higher data diversity and maintain high-accuracy. If an LLM can be used, such restrictions won't be necessary.

For both off-device and on-device modes, we rely on an in-house built companion Android app to extract screen representations and to perform macro actions. For off-device mode, we utilize AndroidEnv~\cite{android_env} to communicate between the companion app and our learning environment. For on-device mode, all the models interact with the companion app directly. 

The neural networks are agnostic to whether the Android accessibility tree or screen understanding techniques are used to produce screen representations. We include demonstrations using both data sources in the same pool of training samples. Both approaches have their limitations. There are icons that are unrecognizable by the icon detectors of screen understanding models and the output of text recognizer may contain errors. On the other hand, visible UI elements may be absent in the corresponding accessibility tree if the app contains Web views, Canvas, etc. 

\subsection{\sys Console}
\label{sec_appendix:ui_nav_console}

To collect demonstrations, we have developed a dedicated application, the \sys Console, that can be seen in the right-half side of the screenshots in Fig.~\ref{fig:send_email}--\ref{fig:notification_dot}. At each step of a demonstration, a user specifies a macro action, including action type, referenced element, and action argument (if any), and then requests execution of the action.

It is typically less effort to complete a task using the \sys Console than directly manipulating the device. For example, entering text using the console takes at most four clicks (clicking the target element, opening the drop-down list of candidate texts, selecting the text to input, clicking the focus\_and\_type button), while manipulating a real device requires keying-in individual characters. The \sys Console also exposes system APIs, such as opening an app through intents, that are not available through the actual device. While using the console may encourage users to complete a task in a way that is different than how they might do through a native interface, the main goal of a trained agent is to successfully complete tasks. Whether it behaves like a human is less important.

In the \sys workflow, new human demonstrations are collected only in scenarios where the current version of the agent or the referee make errors. The demonstration collection interface is integrated with the agent and referee. At each step, the agent's choice of an action and its optional argument are assigned to the internal states and are visualized on the GUI. It is not uncommon that an agent produces correct outputs for unseen scenarios due to the neural networks' capability of generalization. In such cases, a demonstrator simply proceeds with a single click to the next step, thus avoiding the effort of manually specifying the action parameters. Error-driven demonstration collection significantly reduces human effort as well as the number of training samples, which ultimately leads to lower training times.

\subsection{Model training details}
\label{sec:training_details}
\paragraph{Training the agent model.}
For the agent model, demo augmentation happens dynamically with a 1\% probability for a sample to remain unchanged. The model is optimized by an Adam optimizer with a fixed learning rate of 1e-3. Initially a training runs up to 100,000 samples and can be terminated earlier if the test accuracy stabilizes. If new demonstrations are added, the agent will be trained with additional 20,000 samples. It is trained on CPU or GPU with a batch size of 256.

\paragraph{Training the referee model.}
For the referee model, each demonstration is augmented to 10 samples at a pre-possessing stage. The model is optimized by an Adam optimizer with a fixed learning rate of 1e-3. A training takes up to 30 epochs and can be terminated earlier if the test accuracy stabilizes. It is trained on CPU or GPU with a batch size of 128.

\subsection{Case study of agent capabilities}
\label{sec:case-studies}
In the following figures we report screenshots and the associated \sys console. The large arrows in purple are drawn on the screenshots to highlight interesting areas. In the console it is the annotated screen, where UI elements are identified using blue and green boxes. An element highlighted by a red box indicates that it is selected to receive the next action.

\begin{figure*}
    \centering
    \subfloat[]{\includegraphics[width=0.48\linewidth]{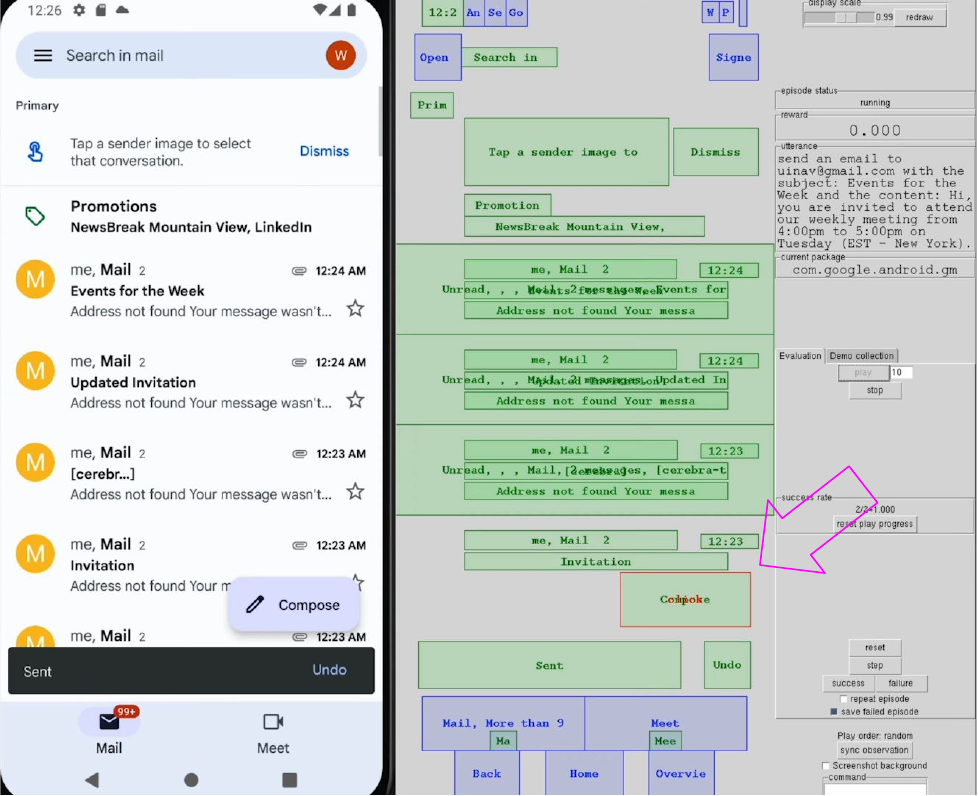}}
    \subfloat[]{\includegraphics[width=0.48\linewidth]{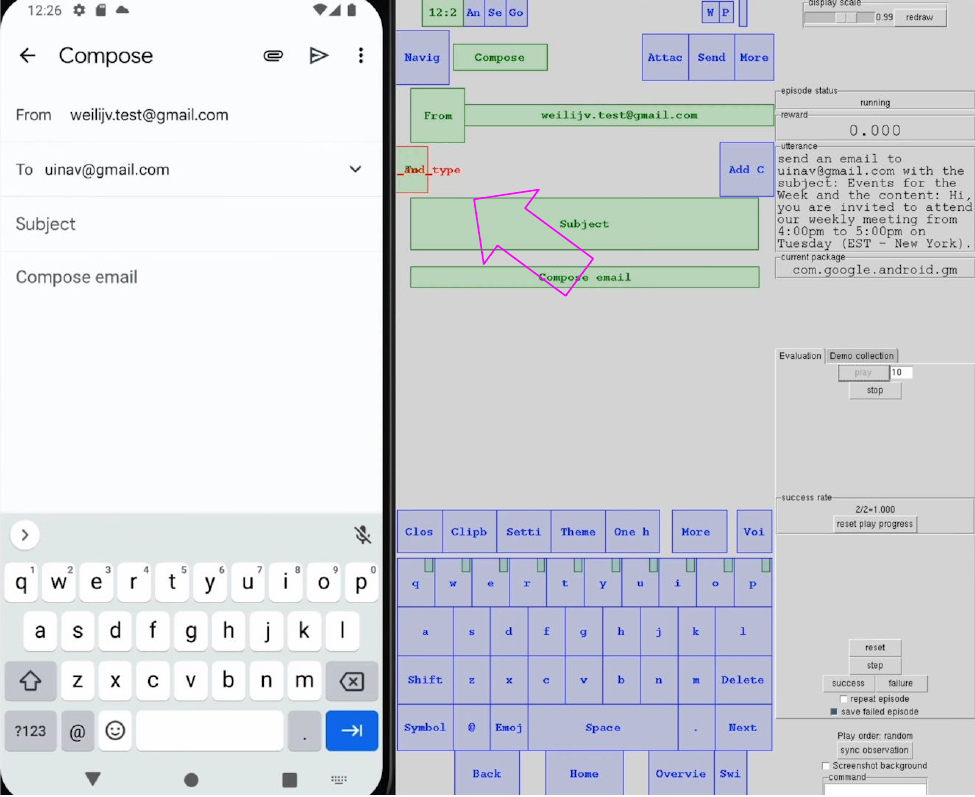}}\\
    \subfloat[]{\includegraphics[width=0.48\linewidth]{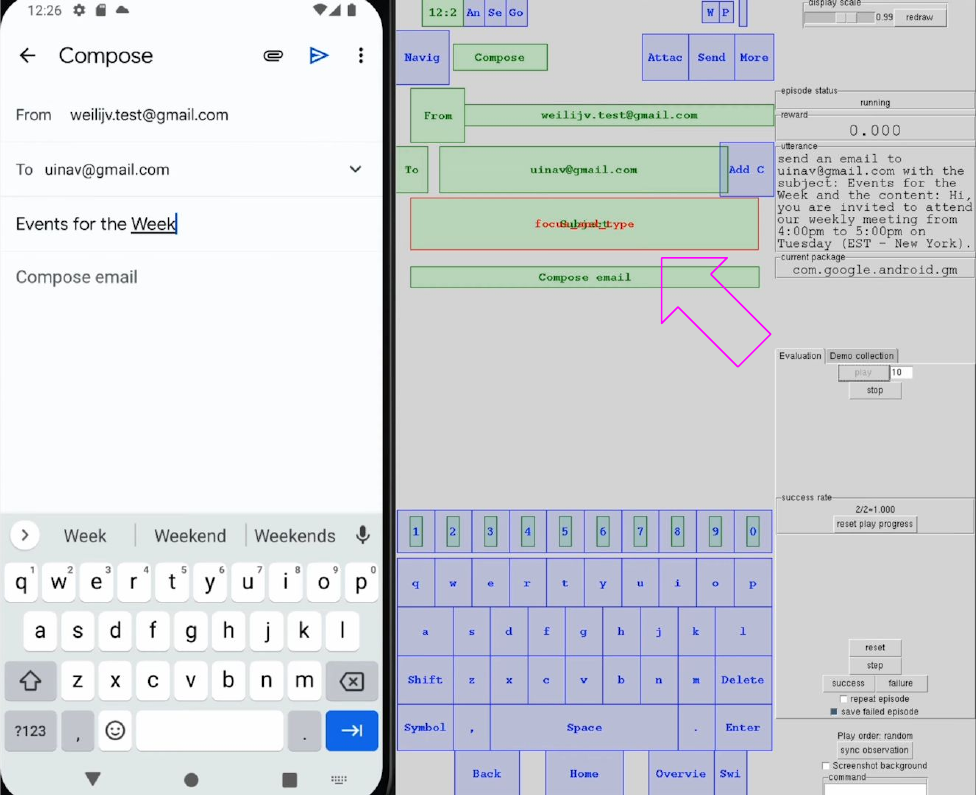}}
    \subfloat[]{\includegraphics[width=0.48\linewidth]{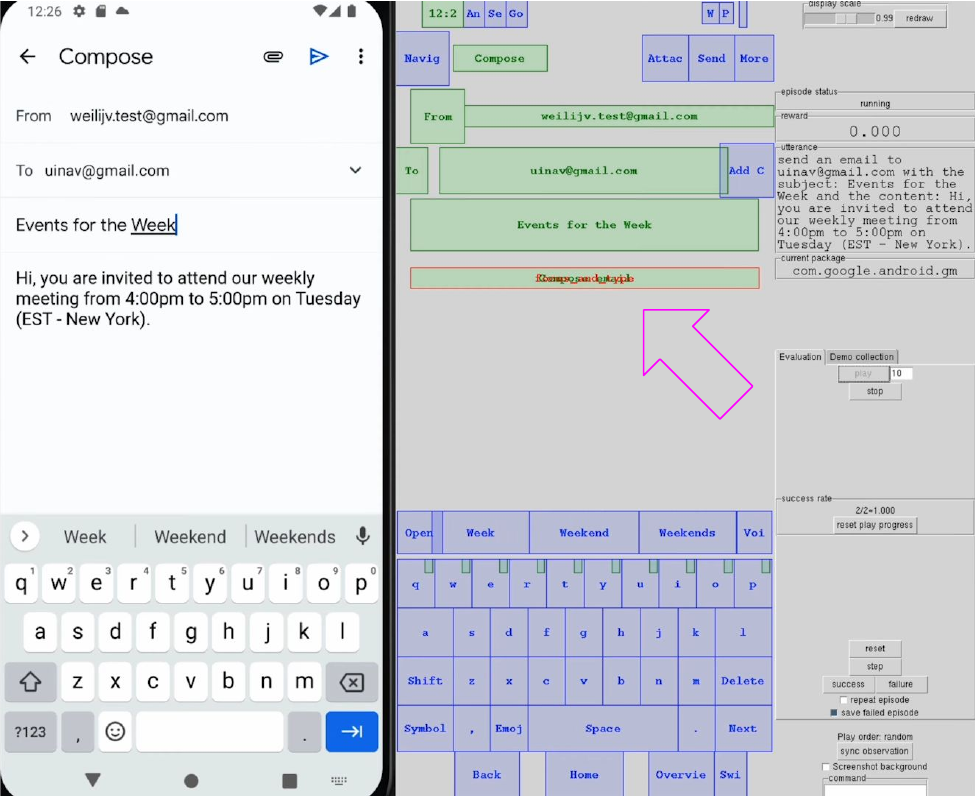}}
    \caption{The \sys agent sends an email: (a) Clicks the compose button; (b) Types the email address; (c) Types the subject; (d) Types the email content. The action of clicking the send button is not shown due to space limitation.} 
    \label{fig:send_email}
\end{figure*}
\begin{figure*}
    \centering
    \subfloat[]{\includegraphics[width=0.48\linewidth]{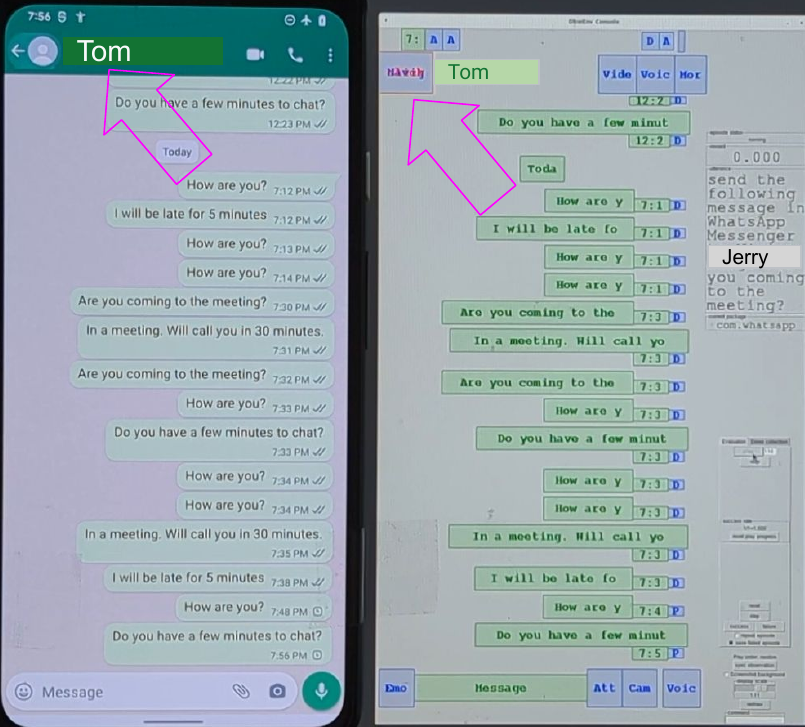}}
    \subfloat[]{\includegraphics[width=0.48\linewidth]{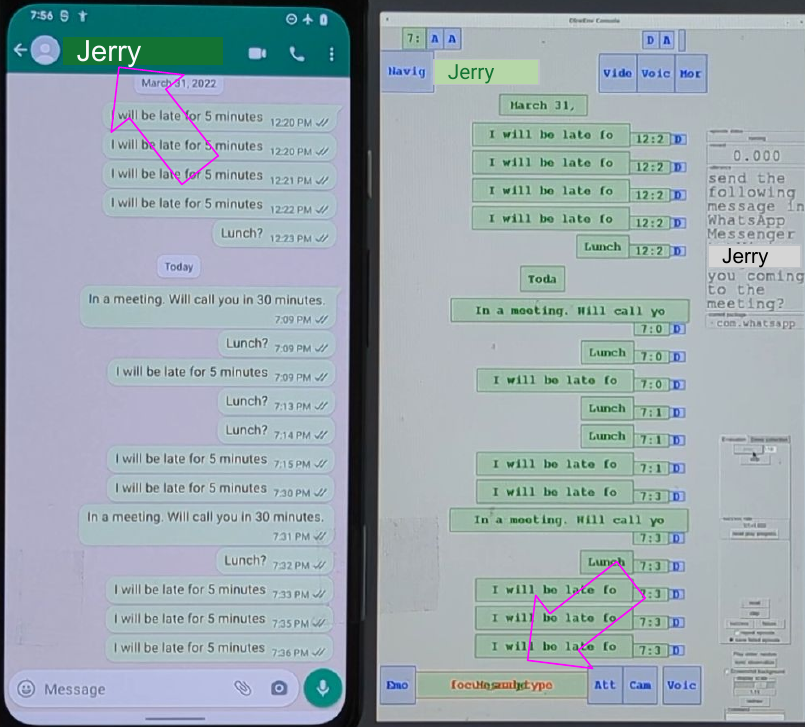}}
    \caption{Two cases of an agent sending a message. The task description is ``send the following message in WhatsApp Messenger to Jerry: Are you coming to the meeting?''. (a) In the message view to a different recipient from the one in the utterance; (b) In the message view of the same recipient as the one in the  utterance.}
    \label{fig:send_message}
\end{figure*}
\begin{figure*}
    \centering
    \subfloat[]{\includegraphics[width=0.5\linewidth]{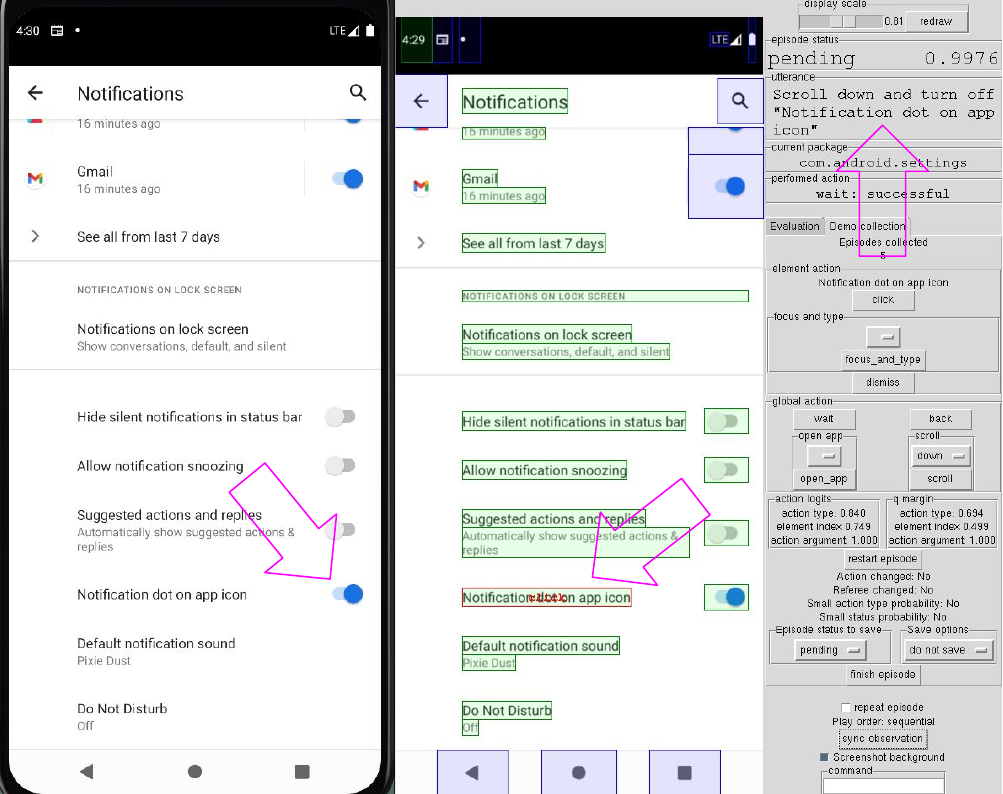}}
    \subfloat[]{\includegraphics[width=0.5\linewidth]{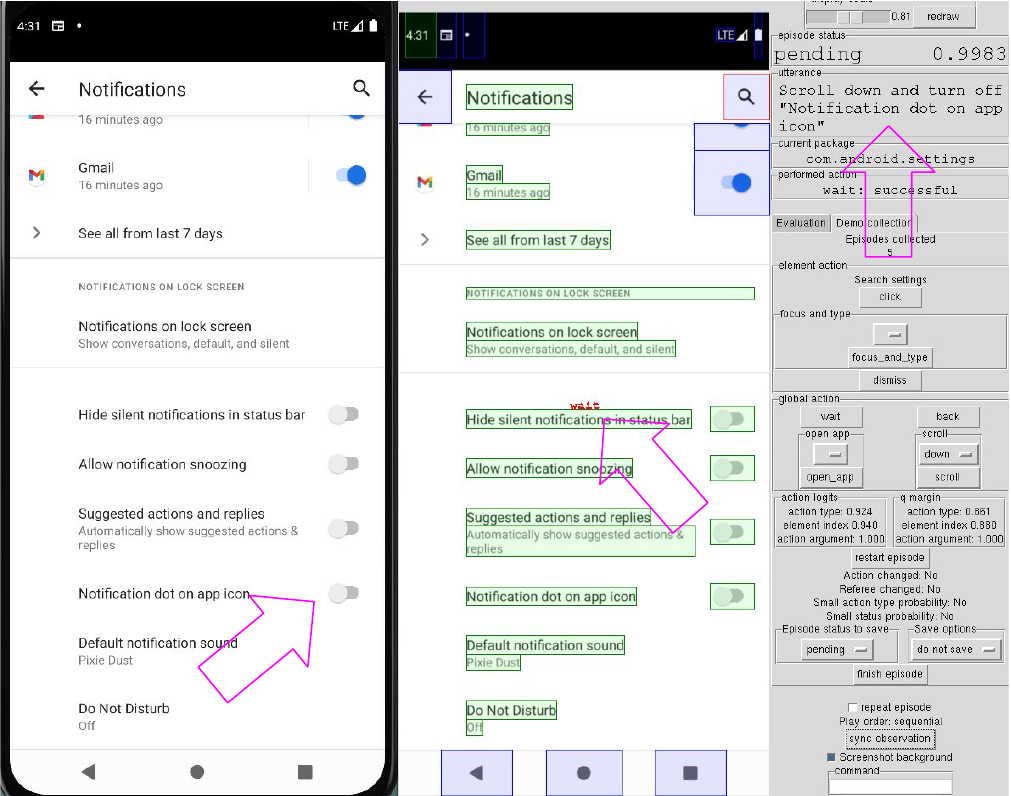}}
    \caption{An agent selects an action to turn off notification dot (a) when the switch is on, and (b) when the switch is already off. The texts in red (click in a) and wait in b)) are the actions selected by the agent.}
    \label{fig:notification_dot}
\end{figure*}

\paragraph{Sending an email with multiple text inputs.}

Fig.~\ref{fig:send_email} shows the image sequence of a \sys agent completing the ``send email'' task. The task utterance is ``send an email to uinav@gmail.com with the subject: Events for the Week and the content: Hi, you are invited to attend our weekly meeting from 4:00pm to 5:00pm on Tuesday (EST - New York)''.

\paragraph{Sending a message to the correct recipient.}

Fig.~\ref{fig:send_message} compares two cases of an agent sending messages. The images are deliberately modified to hide the real names of the recipients. Both (a) and (b) are in the message view of the app but of different recipients, Tom in (a) and Jerry in (b), while the utterance specifies the recipient to be Jerry. The agent correctly recognizes the difference and selects the correct action for both cases: pressing the back button at the top left for (a) and typing the content of the message at the bottom for (b). Note that it is the title bar that contains the information on the current recipient. We believe that it is due to the self-attention of the Transformer encoder that the agent learns whether the text of the title bar matches the recipient is a critical signal in these states.

\paragraph{Understanding the relationship between text label and switch.} 

Fig.~\ref{fig:notification_dot} shows how the \sys agent selects actions to turn off notification dot in two cases: (a) when the switch is on and the agent selects the action to click the text label of "Notification dot on app icon", and (b) when the switch is already off and the agent chooses to wait for the referee to terminate the task. Note that the text label of "Notification dot on app icon" and its switch are independent UI elements in the screen representation, and there are multiple switches on the screen with identical attributes except for their positions and states. The agent learns their relationship probably by the relative positions (horizontally aligned).

\subsection{Apps and websites used in data collection} \label{sec:apps_websites}

The full list of Android apps and websites that are used in our data collection is as follows:

Facebook Messenger,
TikTok,
Instagram,
WhatsApp,
Amazon Shopping,
Facebook,
Walmart,
Spotify,
Pandora,
Amazon Prime Video,
Google Play Games,
Wish,
Pinterest,
Google Messages,
Target,
Poshmark,
Waze,
Twitter,
Wayfair,
google.com,
Google Play Store,
Seamless,
YouTube,
Reddit,
Ebay,
Etsy,
Soundcloud,
Tasty,
Gmail,
Contacts,
Android Auto,
YouTube Music,
Snapchat,
Tubi TV,
Shop,
News Break,
Cash App,
Pluto TV,
Uber,
Burger King,
Roku,
Amazon Alexa,
Life 360,
HBONow,
ESPN,
iHeartRadio,
Nike,
Amazon Photos,
Letgo,
Walmart Grocery,
Weather App,
Google News,
Files,
Home Screen,
Google Docs,
DoorDash,
Google Photos,
AirBnB,
AliExpress,
Amazon Music,
Apple Music,
Audible,
Chewy,
Chik Fil A,
Costco,
Dollar General,
Google Drive,
Dunkin Donuts,
Google Earth,
Emoji Home,
Family Dollar,
wikipedia on firefox,
Food Network,
GroupMe,
Groupon,
GrubHub,
Instacart,
KeepNotes,
King James Version,
Kroger,
Likee,
LinkedIn,
fb Lite,
Lyft,
Maps,
OfferUp,
Phone,
Pixaloop,
Scanner,
SHEIN,
Skype,
SmartNews,
Starbucks,
thredUp,
Ticket Master,
Walgreen's,
Yahoo Mail,
Yelp,
YouTube Kids,
Zedge,
Zelle,
Zillow,
wikipedia.org,
youtube.com,
yahoo.com,
facebook.com,
live.com,
reddit.com,
bing.com,
linkedin.com,
Sam's Club,
discord,
GoodRx,
Outlook,
Breaking US News,
Lucky Go,
CNN,
Postmates,
Transit,
Sephora,
target.com,
twitter.com,
irs.gov,
craigslist.org,
homedepot.com,
Recipes Home,
Zillow, and
Dialer.


\end{document}